\RequirePackage{fix-cm}
\documentclass[conference]{IEEEtran}
 \usepackage{eso-pic}
 \usepackage{color}
 \definecolor{DisclaimerGray}{gray}{0.92}
 
 \AddToShipoutPicture*{\put(45,760){\colorbox{DisclaimerGray}{\parbox{\textwidth}
 {\footnotesize \copyright IEEE, 2018. This is the author's version of the work. Personal use of this material is permitted. However, permission to reprint/republish this material for advertising or promotional purpose or for creating new collective works for resale or redistribution to servers or lists, or to reuse any copyrighted component of this work in other works must be obtained from the copyright holder. The definite version is published at IEEE ENERGYCON 2018.}}}}

\usepackage[noadjust]{cite}
\usepackage{graphicx}
\usepackage{url}
\urldef{\test}{\url}{http://de.wikipedia.org/wiki/Hilfe:TeX}
\usepackage{siunitx}
\usepackage{booktabs}
\usepackage{cleveref}
\usepackage{datetime}
\usepackage{etoolbox}
\makeatletter
\patchcmd{\@makecaption}
  {\scshape}
  {}
  {}
  {}
\makeatletter
\patchcmd{\@makecaption}
  {\\}
  {:\ }
  {}
  {}
\makeatother
\def\tablename{Tab.}
\renewcommand{\thetable}{\arabic{table}}
\usepackage{notoccite}
\begin{document}
\title{Efficiency of Photovoltaic Systems\\
in Mountainous Areas
}


\author{\IEEEauthorblockN{Sri Rama Phanindra Chitturi \hspace{0.5cm} Ekanki Sharma \hspace{0.5cm} Wilfried Elmenreich}
\IEEEauthorblockA{Institute of Networked and Embedded Systems/Lakeside Labs\\
Alpen-Adria-Universit\"{a}t, Klagenfurt, Austria\\
Schittur@edu.aau.at \hspace{0.5cm} Ekanki.Sharma@aau.at \hspace{0.5cm}  Wilfried.Elmenreich@aau.at}}




\date{Received: date / Accepted: date}

\IEEEoverridecommandlockouts
\IEEEpubid{\makebox[\columnwidth]{978-1-5386-3669-5/18/\$31.00~
\copyright2018
IEEE \hfill} \hspace{\columnsep}\makebox[\columnwidth]{ }} 

\maketitle

\begin{abstract}
Photovoltaic (PV) systems have received much attention in recent years due to their ability of efficiently converting solar power into electricity, which offers important benefits to the environment.
PV systems in regions with high solar irradiation can produce a higher output but the temperature affects their performance.
This paper presents a study on the effect of cold climate at high altitude on the PV system output.
We report a comparative case study, which presents measurement results at two distinct sites, one at a height of 612 meters and another one at a mountain site at a height of 1764 meters.
This case study applies the maximum power point tracking (MPPT) technique in order to determine maximum power from the PV panel at different azimuth and altitude angles.
We used an Arduino system to measure and display the attributes of the PV system.
The measurement results indicate an increased efficiency of 42\% for PV systems at higher altitude. \\
\begin{IEEEkeywords}
Photovoltaics (PV), high altitudes, maximum power point tracking (MPPT)
\end{IEEEkeywords}

\end{abstract}
\section{Introduction}
\label{intro}
Energy is an important global issue as currently most of the energy comes from fossil fuels.
If fossil fuel resources will continue to be consumed at the current rate it will be completely dissolved in 100-150 years~\cite{sasaki2011wireless}.
Furthermore, its consumption raises serious environmental concerns as this increases carbon dioxide $(CO_{2})$ and greenhouse gases in the atmosphere.
A possible solution to this problem is to utilize renewable energy resources.
Among all the renewable energy sources, solar power is the one of most promising and free of operational cost energy source~\cite{sasaki2014s}.
PV cells are a promising technology to utilize solar power and convert it directly to electricity. 


In general, solar power generation works better in areas with large solar irradiation. Studies have shown the potential in tropical~\cite{farhoodnea:15} or desertic~\cite{kazem:14} environments.
However, PV systems are effected by temperature. When sunlight strikes the PV module some of the energy is converted to electricity, but most of the energy just heats up the module \cite{osman2015comparative}.
This heat is dissipated to the environment, where the dissipation rate depends on the ambient temperature.
Thus, in hot climate the module becomes hotter and as the temperature increases above \SI{25}{\degree}, the module voltage decreases by approximately \SI{0,5}{\percent} per \si{\celsius}~\cite{1_microgridinstitute_2017}.
In a cold environment, the electrons in the cell are less agitated.
As a result, the PV cells operate more efficiently at low temperatures and hence this improves the power output.

PV systems on mountains have potential for improvements over PV systems in a valley, as the environment on mountains offers benefits such as less fog, cool temperature and low land price.
The goal of this paper is to present measurements for assessing and validating the potential of PV system on mountains in Austria.

The paper is organized as follows:
\Cref{sec:BG} addresses state-of-the-art and related work on solar power generation at high altitude.
The effects of photovoltaic output are discussed in \Cref{sec:SI}.
\Cref{sec:MPP} explains the maximum power point tracking method that was used in this research.
\Cref{sec:ES} depicts the experimental setup and presents the results followed by \Cref{sec:CO} which concludes the paper. 
\raggedbottom
\section{Background}\label{sec:BG}
There are several studies present in the literature about solar power generation using PV panels, but the efficiency of PV systems is strongly influenced by weather conditions.
Many researches are dedicated to increase the efficiency of solar cells for future applications.

In order to utilize the solar energy available in the high atmosphere it is necessary to have a high altitude platform to support appropriate devices (e.g., PV devices).

There are many different approaches proposed to generate solar power in high altitudes.
In $1970$, Glaser proposed a concept~\cite{glaser1974feasibility} that collects  solar energy using a large satellite (which would continuously capture the full strength of solar radiation) and transmits this energy to the ground by using microwave radiation.
The receiving station then converts the microwave radiation into electric energy which would be available to the users.
The original concept has been continuously revised with advanced technological considerations and the research on this concept is still going on \cite{sasaki2013microwave}.

Researchers also proposed to collect the solar energy using a high altitude aerostatic platform~\cite{aglietti2008solar,aglietti2008aerostat, aglietti2008high}.
This approach overcomes the weather condition related issues.
The platform will be above the clouds except for varying extreme weather conditions.
Since the platform is above the troposphere, the sun rays will travel through less air mass compared to an installation at ground level.

Some researchers proposed harvesting energy at high altitudes by utilizing the strong winds existing in high atmosphere~\cite{roberts2007harnessing} by using flying electrical generators (FEG).
These are basically wind turbines collecting wind power at altitudes from few hundred meters to \SI{10}{\kilo\meter}.
Studies reported in \cite{holzworth1981high, hopfield1971tropospheric} state the use of weather balloons to monitor the output voltage of PV panels at high altitudes.

The studies in \cite{royer1999photovoltaics,panjwani2014effect} suggest the coldest geographical locations on the earth to have the best solar power generation potential when using PV panels, since their efficiency increases with low temperatures.

All these studies state that harnessing energy at high altitudes can generate more power compared to installations at sea level.

\section{Physical, Economic and Legal Aspects}\label{sec:SI}

\subsection{Influences on Solar Irradiance}

The output of PV systems is sensitive to weather conditions, as it depends on the strength of solar radiance striking the PV system.
The amount of the solar irradiance at a given meteorological and geographical location depends on the weather data such as sunshine hours, relative humidity, maximum and minimum temperatures, and cloud coverage.

\begin{figure}[h]
\centering
  \includegraphics[scale = .4] {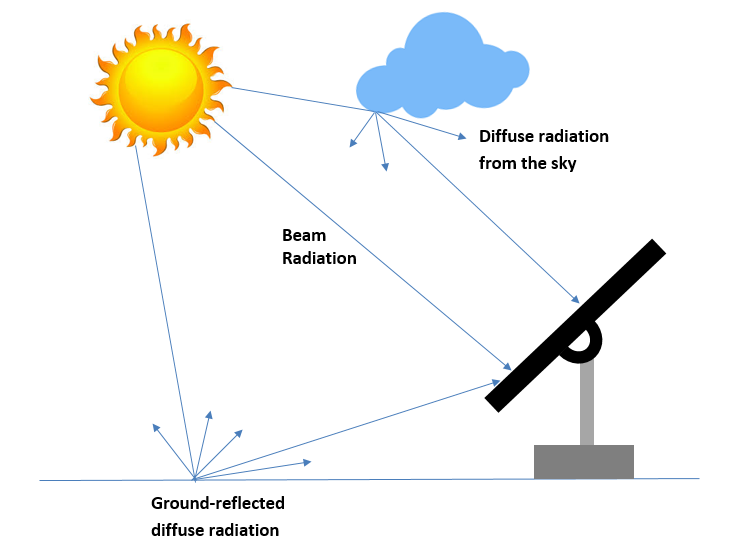}\vspace{-0.5em}
\caption{Global radiation}
\label{fig:1}       
\end{figure}

Solar radiation takes a long way to reach the earth`s surface.  Hence, while modeling the solar radiation, atmospheric particles and radiation models have to be considered.
The radiation that neither reflects nor scatters by molecules in the atmosphere and  reaches the surface directly is known as direct radiation.
The radiation that get scattered from all directions except the disc of the sun is known as diffuse radiation. The sunlight that arrives at the surface after getting reflected (bounced-off) from the direction of ground and reaches an inclined plane is known
as ground-reflected radiation.
The sum of these radiations ~\cite{khatib2016modeling} is known as the global radiation as depicted in \Cref{fig:1}.
\vspace{-0.5em}

\begin{figure}[h]
\begin{center}
  \includegraphics[scale = 0.39] {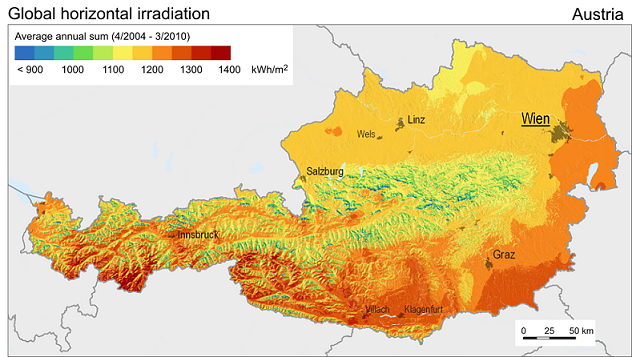}
 \end{center} \vspace{-1em}
\caption{Global solar irradiation map, Austria~\cite{solar}}
\label{fig:2}       
\end{figure}

Global Horizontal Irradiation (GHI) is a reference parameter for calculating the solar radiation on tilted surfaces and for the comparison of climate zones.
\Cref{fig:2} shows the global horizontal irradiation map of Austria.

Sunshine duration is the most useful parameter for estimating the global solar radiation~\cite{udo2002contribution, el2005sunshine, jain1986global, bakirci2009correlations}.
They are a climatological indicator usually expressed as an annual average value.
They are defined as the duration of sunshine in a given period of time (a day or a year) for a particular region on Earth.
The sunshine duration can be reliably and easily measured, as the data is available from decades of weather recordings.
In 1924, {\AA}ngstr{\"o}m proposed a basic model.
He stated that, for estimating the monthly average global solar radiation on horizontal surface, sunshine duration data can be used~\cite{angstrom1924solar}.
\Cref{fig:2} shows the solar irradiation map that provides an annual average sum of concentrating solar power. 
These maps provide a visual presentation of the solar resources and are often used to acquire the ability of solar power generation in a specific region.
Hence they can be used to visually identify the areas rich in solar resources. 
\begin{figure}[h]
\begin{center}
  \includegraphics[scale = 0.65] {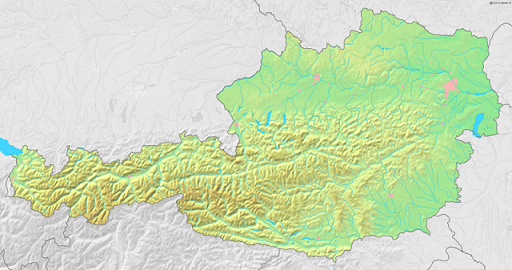}
   \end{center}  \vspace{-1em}
\caption{Topographical map, Austria\cite{topography}}
\label{fig:3}       
\end{figure}

When comparing the global horizontal irradiation map of Austria to a topological map of the same area (see \Cref{fig:3}) we see a correlation between mountainous areas and high global horizontal irradiation.

A major part of Austria is occupied with Alps mountains and solar radiation potential is shown to be high in many of these regions. According to the latest studies harvesting solar power at high altitudes is more efficient than at sea level~\cite{aglietti2008solar}.
The solar irradiation and topographical maps state that the south-west region of Austria has more solar irradiation potential and has lot of mountainous regions.
This validates that at high altitudes in mountainous area, as the slope increases, we get more irradiation (direct radiation) and less diffusion.
Hence at higher altitudes, the availability of full solar radiation allows to form an efficient PV system as compared to ground mounted PV systems. 

\subsection{Economic aspects}

A lot of space is available in the mountain region and cost of lands is also expected to be cheap (Hushak and Sadr~\cite{hushak1979} found that the land price for parcels being far away from urban center and highways is generally lower; Turner, Newton and Dennis~\cite{turner1991} show that there land price is negatively correlated with the portion of parcel with $>$ 15\% slope).

A considerable number of places in the Austrian alps have roads and power lines nearby, especially in the vicinity of power plants and ski resorts. This allows for an injection of the generated power into the grid without further infrastructure. On the other hand, laws regarding protection of the landscape have to be considered, since a large photovoltaic area on a mountain above the treeline would be widely visible. National park areas explicitly forbid  interventions into the landscape by law ~\cite{K-NBG:2013,S-NPG:2015}.

\subsection{Temperature effects}
\begin{figure} [h]
\vspace{-1em}
\begin{center} 
  \includegraphics[width= 7.5cm] {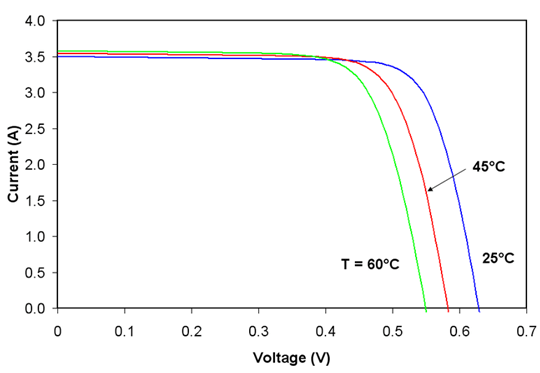}
   \end{center} \vspace{-1.5em}
\caption{ Influence of temperature on $V_{oc}$~\cite{solarcells}}
\label{fig:4}       
\end{figure}

Photovoltaic cells are sensitive to temperature like all other semiconductor devices.
As the temperature increases, the energy of the electrons in the material also increases which in turn reduces the band gap of the semiconductor.
This agitation in electrons, generates slightly more current but less voltage. As a result, the potential difference (voltage) affects the open circuit voltage ($V_{oc}$) as well as the maximum power point and the overall power is reduced.
This influence of temperature on ($V_{oc}$) is illustrated in \Cref{fig:4}.

The current-voltage (I-V) curve ranges from zero current at the open-circuit voltage $V_{oc}$, to the short-circuit current $I_{SC}$ at zero voltage.
The increase in current with increase in temperature is very little and can be ignored.
But, the effect of the increase in temperature on the voltage is noteworthy.
As it is shown in \Cref{fig:4}, there is a linear relationship between voltage and temperature till the temperature reaches \SI{25}{\degree}~\cite{1_microgridinstitute_2017}, but after this the voltage start to decrease with further increase in temperature.

On mountains, the ambient temperature is in general decreasing with height. Although temperature inversion effects are possible in such regions as well, they are considered to have a lesser effect on solar power, since such effects are typically happen in Winter season (when general temperature is low) and during night time.

\section{Maximum power point tracking}\label{sec:MPP}
PV panels exhibit nonlinear I-V and power-voltage (P-V) characteristics which depend on solar irradiation and solar cell temperature~\cite{fesharaki2011effect}.
In order to continuously generate maximum power from the PV panels, they must operate at their maximum power point (MPP) at different weather conditions.
To serve this purpose, maximum power point tracking (MPPT) algorithms are employed to extract the maximum power from the photovoltaic arrays and to increase it's efficiency~\cite{gil2016maximum}.
To measure this maximum power point, an MPP tracking method is used in this work.

To track the maximum power point the microcontroller steps through a number of different load impedances and records the highest power value.
To reduce the effect of noise, each power estimation, done by metering voltage and current, is based on the median value from three consecutive measurements.
A measurement cycle consists of 255 steps, where each step corresponds to a different load value.
For each step, a given PWM value is generated, which is smoothed by a low pass filter and fed into a transistor.
Due to the effects of the low-pass filter and the nonlinearities of the transistor, the particular steps corresponding to different granularities, which is overcome by choosing a sufficiently fine step-size.
\section{Experimental setup}\label{sec:ES}
\label{mes}
\begin{figure} [h]
 \begin{center} 
  \includegraphics[width= 8.5cm] {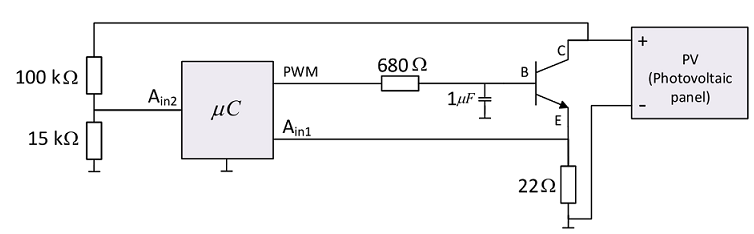}
   \end{center} \vspace{-1em}
\caption{Measurement setup block diagram}
\label{fig:5}       
\end{figure}

\begin{figure} [h]
\vspace{-0.3em}
 \begin{center} 
  \includegraphics[width= 7.5cm] {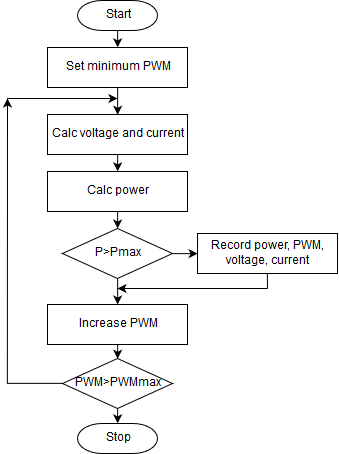}
   \end{center} \vspace{-1em}
\caption{MPPT algorithm flowchart}
\vspace{-0.2em}
\label{fig:6}       
\end{figure}

We implemented a simple measurement setup as shown in \Cref{fig:5}. It consists of a photovoltaic panel, a microcontroller, voltage and current sensing elements.
The objective is to measure the maximum power point (MPP), the current (I), and the voltage (V) of the PV panel.
The power (P) is the combination of ($V\times I$) and measured in watts.
For the proposed measurement setup an Arduino UNO board is used.
The board comprises an Atmel Atmega328 microcontroller, which provides sufficient computational power for performing an MPPT algorithm.
The voltage and current sensors sense the output voltage and output current respectively.
These analog signals are measured via the analog channels $A_{in1}$ and $A_{in2}$ on the Arduino board respectively.
The Atmega328 microcontroller has a built in analog-to-digital (ADC) converter module, so the analog output of the panel voltage and current voltage drop are directly connected to the ADC pins of the microcontroller for further processing.
The MPPT algorithm shown in \Cref{fig:6} is implemented on the Atmega328 microcontroller.

The output of the microcontroller generates a pulse width modulated signal.
Our design uses different duty cycles to adjust the impedance of the photovoltaic panel to reach the MPP.
The PWM (pin 9) increases or decreases the duty cycle, earlier set with a quantized value of 255 for maximum (\SI{100}{\percent}) duty cycle and 0 for minimum (\SI{0}{\percent}) duty cycle respectively.
The switching pulse is necessary to control the operating point of the BJT (BC547B: NPN Bipolar transistor).
The higher the duty cycle, the more current passes from the input to the output. 

The microcontroller requires (panel voltage and current) feedback signals from the PV panel and adjusts the PWM signal to drive the BJT in order to transfer the maximum power.
We have measured the voltage and current for declination angles between \SI{0}{\degree} to \SI{90}{\degree}.

\begin{figure} [h]
\begin{center}
  \includegraphics[width= 8cm] {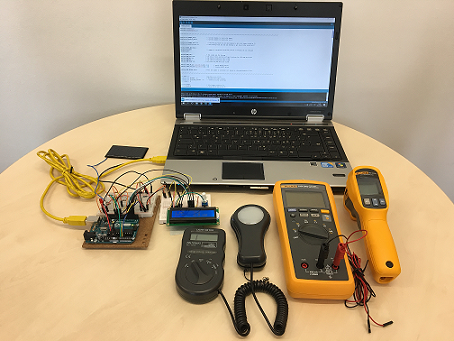}
  \end{center}\vspace{-1em}
\caption{Measurement devices}
\vspace{-1em}
\label{fig:7}       
\end{figure}

The program is written in the "Arduino C++" programming language and was designed with the Arduino software (IDE). The microcontroller was programmend and debugged via a USB-to-serial interface.
A liquid crystal display (LCD) is used to display the parameters (voltage (V), current (I), power (P), duty cycle (DC)).
\Cref{fig:7} depicts the measurement devices used for the experiment.

\begin{figure} [h]
\begin{center}
  \includegraphics[width= 8cm] {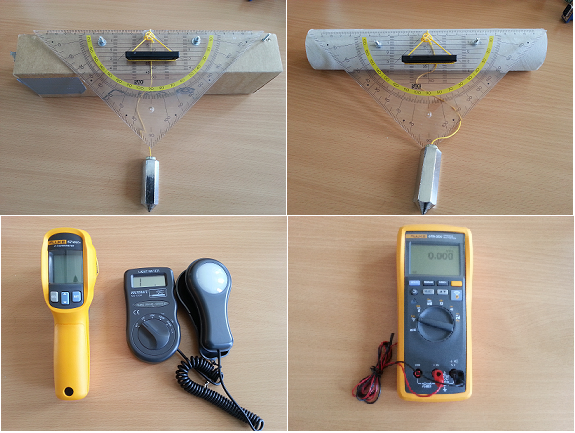}
  \end{center}\vspace{-1em}
\caption{Measurement apparatus}
\label{fig:8}       
\end{figure}

To study the potential of solar energy in high altitudes, we have tested the designed measurement setup in two distinct sites on a sunny day.
We prepared a measurement protocol sheet in which we recorded voltage, current, panel temperature, relative humidity, ambient temperature, and light intensity.
To determine the declination angle we applied a protractor.
The angular interpretation was changed, so that the \SI{90}{\degree} label represents \SI{0}{\degree}.
Angles on the left are signed positive whereas angles on the right are signed negative.
With this the possible angles ranges from \SIrange{90}{-90}{\degree} (counterclockwise).
The power measurement was performed using declination angles between \SI{0}{\degree} to \SI{90}{\degree}.
The light intensity measurements were done using declination angles between \SI{0}{\degree} to \SI{90}{\degree} to \SI{-45}{\degree}.
The apparatus required for this measurements is shown in \Cref{fig:8}.

Their model numbers are provided in \Cref{tab:Apperatus}:

\begin{table}[h]
\centering
\begin{tabular}{lll} 
\toprule
\textbf{APPERATUS} & 
\textbf{MODEL NUMBER} &
\textbf{Usage}\\
\midrule
Photometer	& VOLTCRAFT MS-1300		& Light intensity \\
Thermometer	& FLUKE 62 MAX+			& Temperature \\ 
Hygrometer	& TFA Dostmann/Wertheim & Humidity \\ 
			& Model: 35.1024.54		&\\
Multimeter	& FLUKE CNX 3000		& Voltage, Current \\
Plummet		& SILVERLINE			& Declination angles\\
Protractor	&						& Declination angles\\
\bottomrule
\end{tabular}
\vspace{10pt}
\caption{Measurement Apparatus}\vspace{-0.7cm}
\label{tab:Apperatus}
\end{table}
\vspace{0.35em}

\begin{figure} [h]
\begin{center}
  \includegraphics[width= 8cm] {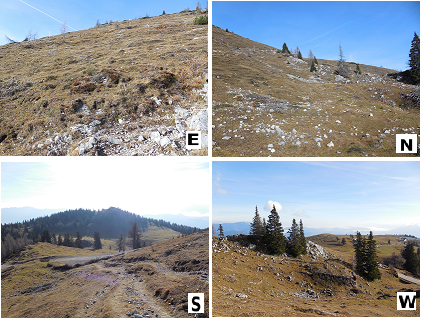}
  \end{center}\vspace{-1em}
\caption{Measurement site at altitude}
\label{fig:9}       
\end{figure}

The first measurement site was "Dobratsch above Rosstratten" at an altitude of \SI{1764}{m}.
The coordinates of the site are $46^\circ35.776'$ North, $13^\circ42.330'$ East.
The view towards the four cardinal directions of the site are shown in \Cref{fig:9}.
The measurements have been taken on the date $23.11.2016$ between 10:40 am and 11:00 am.
At the time of measurement, the weather was sunny and the panel temperature was at \SI{6.7}{\degreeCelsius}.
The ambient temperature was \SI{6}{\celsius} and the relative humidity was \SI{62}{\percent} respectively.

\begin{figure} [h]
\begin{center}
   \includegraphics[width= 8cm] {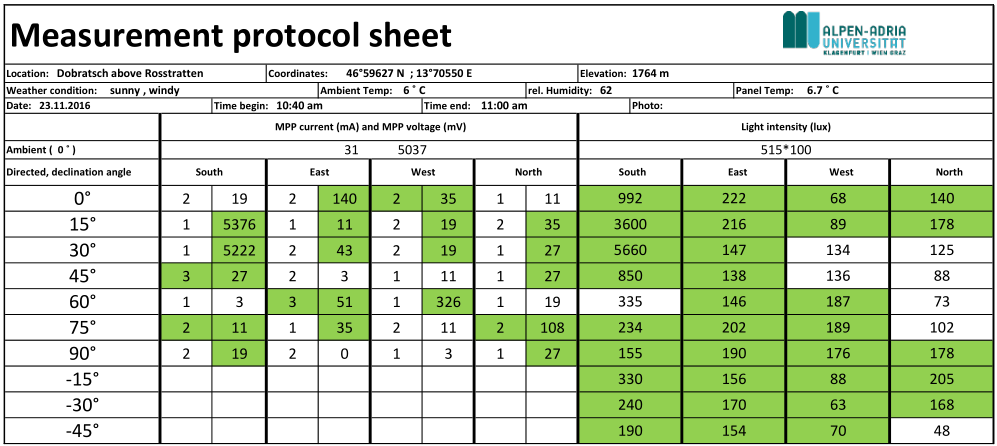}
   \end{center}
   \vspace{-1em}
\caption{Measurement protocol sheet of dobratsch above rosstratten}
\label{fig:10}       
\end{figure}

The measurements on indirect radiation, the ambient power was \SI{156,147}{\milli\watt} and light intensity \SI{515}{\lux}.
For directed radiation, the declination angle \SI{15}{\degree} to \SI{30}{\degree} yielded the maximum power \SI{5,376}{\milli\watt} and light intensity \SI{5660}{\lux} in south. 
The measurement protocol sheet of at high altitude is shown in \Cref{fig:10}. 
The highlighted numbers show, where the output was improved at the measurement site at altitude.

\begin{figure} [h]
\begin{center}
  \includegraphics[width= 8cm] {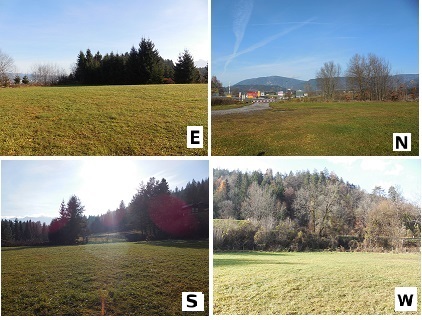}
  \end{center}\vspace{-1em}
\caption{Measurement site at ground level}
\vspace{-1em}
\label{fig:11}       
\end{figure}

The second measurement site was "Beginn Alpenstrasse" at an altitude of \SI{612}{\meter}.
The coordinates of the site are $46^\circ36.121'$ North, $13^\circ48.765'$ East.
The four cardinal directions of the site are shown in \Cref{fig:11}.
The measurements have been taken on the date $23.11.2016$ between 11:37 am and 11:50 am.
At the time of measurement, weather was sunny and the panel temperature was \SI{21,8}{\celsius}.
The ambient temperature was \SI{17}{\celsius} and the relative humidity was \SI{56}{\percent} respectively.

\begin{figure} [h]
\begin{center}
   \includegraphics[width= 8.5cm] {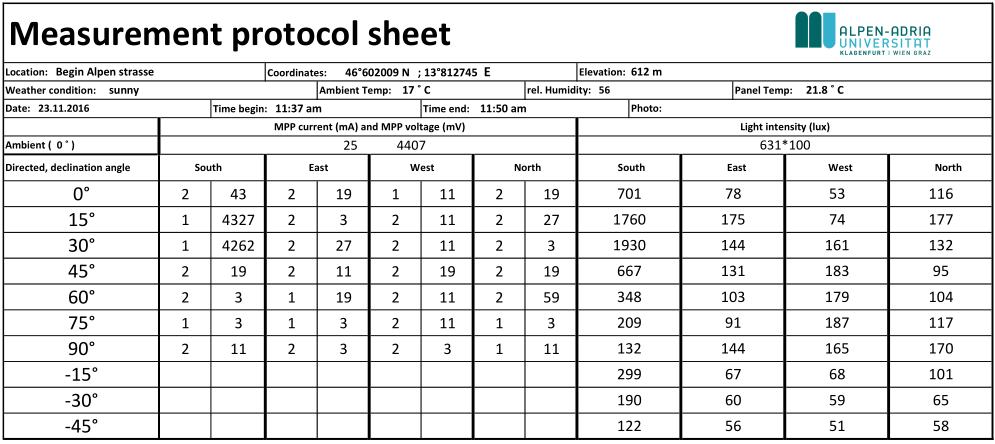}
   \end{center}\vspace{-1em}
\caption{Measurement protocol sheet of begin Alpenstrasse}
\label{fig:12}       
\end{figure}

The measurements on indirect radiation, the ambient power was \SI{110,175}{\milli\watt} and light intensity \SI{631}{\lux}.
For direct radiation, the declination angles between \SI{15}{\degree} to \SI{30}{\degree} yielded the maximum power \SI{4,327}{\milli\watt} and light intensity \SI{1930}{\lux} in south. 
The measurement protocol sheet of ground level is shown in \Cref{fig:12}.

\begin{figure} [h]
\begin{center}
   \includegraphics[width= 8.5cm] {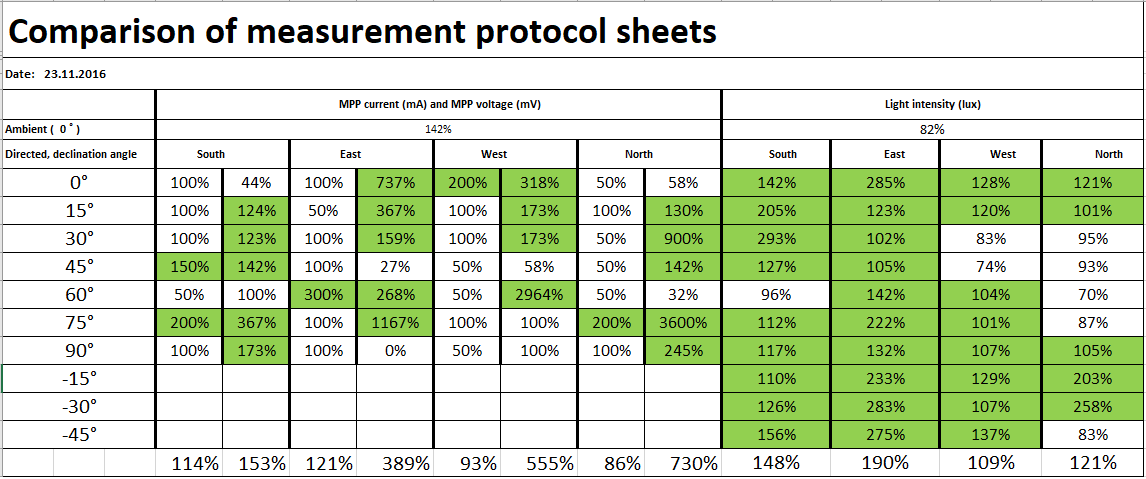}
   \end{center}\vspace{-1em}
\caption{Measurements comparison of two distinct sites, highlighted numbers show, where the output was improved at the measurement site at altitude.}
\vspace{-0.5em}
\label{fig:13}       
\end{figure}

The meteorological data were collected at the sites.
The data values are compared and stored in an Excel sheet for evaluating the improvement of site1 over site2.
The results of voltage, current and light intensity are shown in \Cref{fig:13}.

The measurements on indirect radiation, i.e., the average ambient power was \SI{142}{\percent} and light intensity \SI{82}{\percent} in comparison to the corresponding values at ground level.
A detailed explanation of measurement setup and results can be found in~\cite{chitturi:Thesis:2017}.

\section{Conclusion and Future work}\label{sec:CO}
This paper reported measurement results from subsequent measurements executed on the same day at different altitudes. 
Although the day was very clear, there were significant differences in power output and solar irradiation between the two sites.
The measurements have been taken within a duration of 70 minutes, where the measurements at the lower altitude was closer to noon than the measurements at the higher altitude. 
Despite this, the output of the solar panel was measured to be higher in altitude compared to the ground level. 
Measurements have been done for ambient irradiation as well as for directed irradiation. 
Results show that the power output in the ambient case was \SI{42}{\percent} higher, but interestingly the ambient irradiation was measured to be \SI{18}{\percent} lower at the mountainous measurement site. 
This could be due to measurement variation or different measurement instants.

For the directed measurements the four cardinal directions had been sampled for elevation angles of different multiples of \SI{15}{\degree}. 
In the peak case, which was south at an elevation angle of \SI{15}{\degree}, the mountainous measurement site yielded \SI{24}{\percent} higher power output.

The presented measurements are subject to weather and device variation, but indicate a possible higher potential for PV systems when deployed at sites with a higher altitude. 
In contrast, infrastructure costs for the installation and grid connection have to be considered when designing PV plants at mountainous sites. 
Further issues could be higher maintenance costs due to snow and ice. \newline
The work can be extended further to include a mathematical model to quantify and validate the findings obtained from the experimental setup.   

\section*{Acknowledgments}

We thank Sascha Einspieler for his constructive comments and his valuable writing tips. Furthermore we are grateful to the anonymous reviewers for their careful reading of our manuscript and their  
insightful comments and suggestions.



\bibliographystyle{IEEEtran}
\bibliography{IEEEexample}



\end{document}